\documentclass[twocolumn,showpacs,preprintnumbers,superscriptaddress,prl,amsmath,amssymb,nofootinbib]{revtex4-1}

\usepackage{graphicx}
\usepackage{dcolumn}
\usepackage{bm}
\usepackage{color}
\usepackage{ulem}
\usepackage{gensymb}
\usepackage{braket}
\usepackage{epstopdf}
\usepackage{amsmath}
\usepackage[percent]{overpic}

\begin{document}

\newcommand{\customsection}[1]{\textit{#1.\textemdash}}
\newcommand{\temperaturerange}[3]{#1\,K $\leq$ #2 $\leq$ #3\,K}

\title{Crystal growth and magnetic structure of MnBi$_2$Te$_4$\\}

\author{J.-Q. Yan}
\affiliation{Materials Science and Technology Division, Oak Ridge National Laboratory, Oak Ridge, Tennessee 37831, USA}
\email{yanj@ornl.gov}

\author{Q. Zhang}
\affiliation{Neutron Scattering Division, Oak Ridge National Laboratory, Oak Ridge, Tennessee 37831, USA}

\author{T. Heitmann}
\affiliation{The Missouri Research Reactor, University of Missouri, Columbia, Missouri 65211, USA}

\author{Z. L. Huang}
\affiliation{Department of Physics and Astronomy, Rutgers University, Piscataway, New Jersey 08854, USA}

\author{W. D. Wu}
\affiliation{Department of Physics and Astronomy, Rutgers University, Piscataway, New Jersey 08854, USA}

\author{D. Vaknin}
\affiliation{Ames Laboratory and Department of Physics and Astronomy, Iowa State University, Ames, Iowa 50011, USA}

\author{B. C. Sales}
\affiliation{Materials Science and Technology Division, Oak Ridge National Laboratory, Oak Ridge, Tennessee 37831, USA}

\author{R. J. McQueeney}
\affiliation{Ames Laboratory and Department of Physics and Astronomy, Iowa State University, Ames, Iowa 50011, USA}

\date{\today}

\begin{abstract}
Millimeter-sized MnBi$_2$Te$_4$ single crystals are grown out of a Bi-Te flux and characterized using magnetic, transport, scanning tunneling microscopy and spectroscopy measurements. The magnetic structure of MnBi$_2$Te$_4$ below T$_N$ is determined by powder and single crystal neutron diffraction measurements. Below T$_N$=24\,K, Mn$^{2+}$ moments order ferromagnetically in the \textit{ab}-plane but antiferromagnetically along the crystallographic \textit{c}-axis. The ordered moment is 4.04(13)$\mu_{B}$/Mn at 10\,K and aligned along the crystallographic \textit{c}-axis in an A-type antiferromagnetic order. The electrical resistivity drops upon cooling across T$_N$ or when going across the metamagnetic transition in increasing magnetic fields below T$_N$. A critical scattering effect is observed in the vicinity of T$_N$ in the temperature dependence of thermal conductivity. However, a linear temperature dependence is observed for the thermopower in the temperature range 2\,K$\leq$T$\leq$300\,K without any anomaly around T$_N$. Fine tuning of the magnetism and/or electronic band structure is needed for the proposed topological properties of this compound. The growth protocol reported in this work might be applied to grow high quality crystals where the electronic band structure and magnetism can be finely tuned by chemical substitutions.

\end{abstract}

\maketitle

\section{Introduction}

The intersection of magnetism with topological electronic states has become an exciting area in condensed matter physics.  A variety of exotic quantum states have been predicted to emerge, such as the quantum anomalous Hall effect, Weyl semimetals, and axion insulators, although only a few experimental realizations have been found to date. For example, the introduction of bulk ferromagnetic (FM) order in a topological insulator (TI) has been shown to induce the quantum anomalous Hall effect at very low temperatures ($\approx$10\,mK) in thin films of a TI with dilute magnetic doping, such as (Bi$_{1-y}$Sb$_y$)$_{2-x}$Cr$_x$Te$_3$.\cite{chang2013experimental} In this situation, FM order preserves the bulk electronic gap (a FM insulator) and also gaps the spin-momentum locked Dirac-like surface states, producing dissipationless edge modes in the absence of an applied magnetic field. FM order can also close the bulk gap in a TI through exchange coupling, inducing a gapless Weyl semimetal with topologically protected bulk chiral electronic states.  In both cases, the breaking of time-reversal symmetry by the magnetic order is key to the unusual topological properties.

Another interesting approach is to consider the effect of antiferromagnetic (AFM) order in topological materials.  In this case,  time-reversal symmetry is broken, but the combination of time reversal and a half-lattice translation is not broken, which leads to a Z$_2$ topological classification.  Such AFM-TIs are predicted to host unusual quantum axion electrodynamics at the surface.\cite{mong2010antiferromagnetic} However, it is extremely rare to find the naturally-grown mutilayers where magnetic (either ferromagnetic or antiferromagnetic) and topological phases coexist and intimately couple to each other.

It has recently been proposed that MnBi$_2$Te$_4$ may be the first example of an AFM-TI.\cite{otrokov2017highly,zhang2018topological,otrokov2018prediction}  MnBi$_2$Te$_4$ is based on the Bi$_2$Te$_3$ tetradymite structure common to the well-known topological insulators.  The tetradymite structure is rhombohedral and consists of a van der Waals bonded quintuple-layers with a Te-Bi-Te-Bi-Te sequence.  In MnBi$_2$Te$_4$, an additional Mn-Te layer is inserted, Te-Bi-Te-Mn-Te-Bi-Te, forming a septuple-layer.  Magnetic measurements confirm that the Mn ions adopt a high-spin S=5/2 of a 2+ valence with a large magnetic moment of $\sim$5 $\mu_B$ and also indicate an AFM transition at 24\,K.\cite{otrokov2018prediction} Therefore, MnBi$_2$Te$_4$ offers a unique natural heterostructure of antiferromagnetic planes intergrowing with layers of topological insulators. First-principles calculations, magnetic measurements, and X-ray magnetic circular dichroism measurements predict an A-type magnetic structure with FM hexagonal layers coupled antiferromagnetically along the \textit{c}-axis. However, a confirmation of the magnetic structure by neutron diffraction is still absent possibly due to the difficulty of synthesis of polycrystalline samples and growth of sizable single crystals.\cite{zeugner2018chemical}

In this work, we report the growth of sizable single crystals of MnBi$_2$Te$_4$ out of a Bi-Te flux. The as-grown MnBi$_2$Te$_4$ crystals have an electron concentration of 5.3$\times$10$^{20}$cm$^{-3}$ at room temperature and exhibit antiferromagnetic order at T$_N$=24\,K. Our neutron powder and single crystal diffraction measurements confirm the previously proposed A-type antiferromagnetic order with ferromagnetic planes coupled antiferromagnetically along the \textit{c}-axis. The ordered moment is 4.04(13)$\mu_{B}$/Mn at 10\,K and aligned along the crystallographic \textit{c}-axis. The magnetic order affects both the electrical and thermal conductivity. However, we observed no anomaly in the temperature dependence of thermopower around T$_N$.

\begin{figure} \centering \includegraphics [width = 0.40\textwidth] {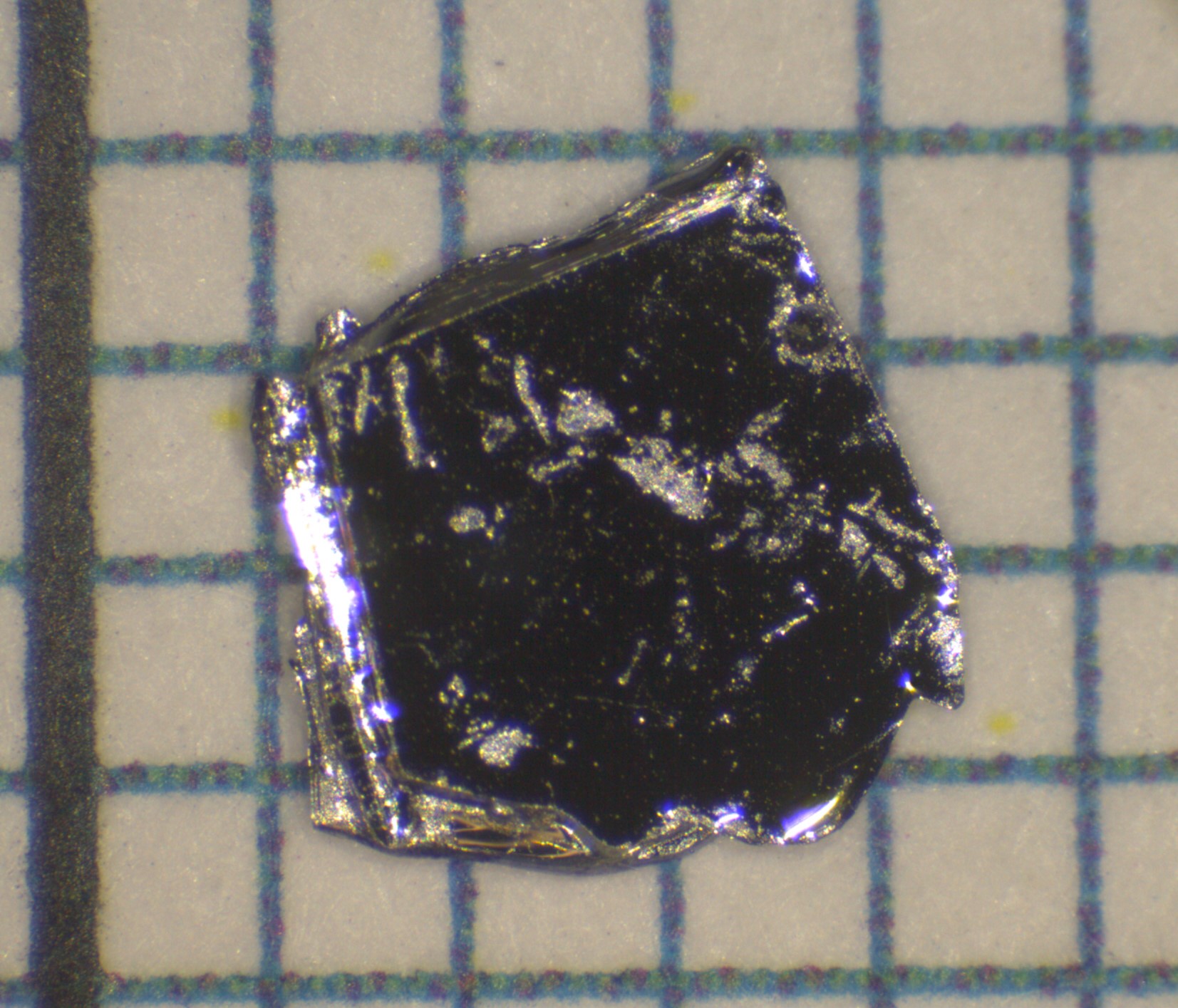}
\caption{(color online) Photograph of a single crystal of MnBi$_2$Te$_4$ on a mm grid.}
\label{picture-1}
\end{figure}

\section{Experimental details}

MnBi$_2$Te$_4$ single crystals were grown by a flux method. Mixtures of Mn (alfa aesar, 99.99\%), Bi pieces (alfa aesar, 99.999\%), and Te shot (alfa aesar, 99.9999\%) in the molar ratio of 1:10:16 (MnTe:Mn$_2$Te$_3$=1:5) were placed in a 2ml alumina growth crucible of a Canfield crucible set,\cite{canfield2016use} and heated to 900$^\circ$C and held for 12 h. After slowly cooling across a $\approx$10 degree window below 600$^\circ$C in two weeks,  the excess flux was removed by centrifugation above the melting temperature of Bi$_2$Te$_3$ (585$^\circ$C). We also tested other ratios of starting materials to find MnTe:Mn$_2$Te$_3$=1:5 gives large crystals and reasonable yield. Crystals produced by this flux method  were typically few mm on a side and often grew in thick, blocklike forms with thicknesses up to 2 mm, but are easily delaminated into thin sheets. Figure\,\ref{picture-1} shows the picture of one crystal. Most crystals are about 20 mg per piece but the thick crystals can be over 50 mg.

We performed elemental analysis on both the as-grown and freshly cleaved surfaces using a Hitachi TM-3000 tabletop electron microscope equipped with a Bruker Quantax 70 energy dispersive x-ray system. The elemental analysis confirmed the Mn:Bi:Te ratio is 14.4:28.6:57.0 in the crystals. Magnetic properties were measured with a Quantum Design (QD) Magnetic Property Measurement System in the temperature range 2.0\,K$\leq$T$\leq$\,300\,K. The temperature and field dependent electrical resistivity data were collected using a 9\,T QD Physical Property Measurement System (PPMS).  One single crystal with dimensions of 1.20\,mm\,$\times$\,0.75\,mm\,$\times$\,7\,mm was selected for the thermal conductivity measurement using the TTO option of 9T PPMS.  Silver epoxy (H20E Epo-Tek) was utilized to provide mechanical and thermal contacts during the thermal transport measurements. The thermal conductivity measurement was performed with the heat flow in the \textit{ab}-plane.

The scanning tunneling microscopy (STM) and spectroscopy (STS) measurements were performed at 4.5 K in an Omicron UHV-LT-STM with a base pressure 1$\times$10$^{-11}$mbar. Electrochemically etched tungsten tips were characterized on Au (111) surface. The MnBi$_2$Te$_4$ single crystals were cleaved in situ at room temperature and then transferred immediately into the cold STM head for measurements. The dI/dV spectra were measured with the standard lock-in technique with a modulation frequency f = 455 Hz and a modulation amplitude Vmod = 20 mV.

Neutron powder diffraction was performed on the time-of-flight (TOF) powder diffractometer,
POWGEN, located at the Spallation Neutron Source at Oak
Ridge National Laboratory. The powder sample used for the neutron diffraction measurements was synthesized by annealing at 585$^\circ$C for a week the homogeneous stoichiometric mixture of the elements quenched from 900$^\circ$C.\cite{zeugner2018chemical} X-ray powder diffraction performed on a PANalytical X'Pert Pro MPD powder X-ray diffractometer using Cu K$_{\alpha1}$ radiation found weak reflections from MnTe$_2$. The room temperature lattice parameters are a=4.3243(2)\,${\AA}$, c=40.888(2)\,${\AA}$, consistent with previous reports.\cite{lee2013crystal,zeugner2018chemical} Magnetic measurement confirmed the polycrystalline sample orders antiferromagnetically at T$_N$=24\,K. Around 2.1 g
powder was loaded in a vanadium container and the POWGEN Automatic Changer was used to access the temperature region of
10$-$300 K.  The data were collected with neutrons
of central wavelengths 1.5\,$\AA{}$. Symmetry allowed magnetic structures are
analyzed using the Bilbao crystallographic server.\cite{aroyo2006bilbao,gallego2012magnetic} All of the neutron diffraction data were analyzed
using the Rietveld refinement program FULLPROF suite.\cite{rodriguez1993recent}

Single-crystal neutron diffraction experiments were carried out on the triple-axis spectrometer (TRIAX) located at the University of Missouri Research Reactor (MURR). The TRIAX measurements utilized an incident energy of $E_i=14.7$ meV using a  pyrolytic graphite (PG) monochromator system and is equiped with an PG analyzer stage. PG filters were placed before and after the second monochromator to reduce higher order contamination in the incident beam achieving a ratio $\frac{I_\lambda}{2}:I_\lambda: \approx 10^{-4}$. The beam divergence was defined by collimators of 60'-60'-40'-80' between the reactor source to monochromator; monochromator to sample; sample to analyzer; and analyzer to detector, respectfully.  A 14 mg MnBi$_2$Te$_4$ crystal was loaded to the cold tip of the Advanced Research Systems closed-cycle refrigerator and cooled to a base temperature of 6.7 K. The sample was mounted with the (1,0,L) plane in the neutron scattering plane, with lattice parameters $a=4.303${\AA} and $c=40.231$ {\AA} at base temperature. The crystal that we examined shows some twinning of two inequivalent domains that are rotated by 120$^\circ$. These twinned grains are identified as Bragg peaks along (H0L) that do not conform to the reflection condition -H+L=3n for the R-3m space group, but are rather (0KL) reflections from the twin which obeys the K+L=3n condition.

\begin{figure} \centering \includegraphics [width = 0.47\textwidth] {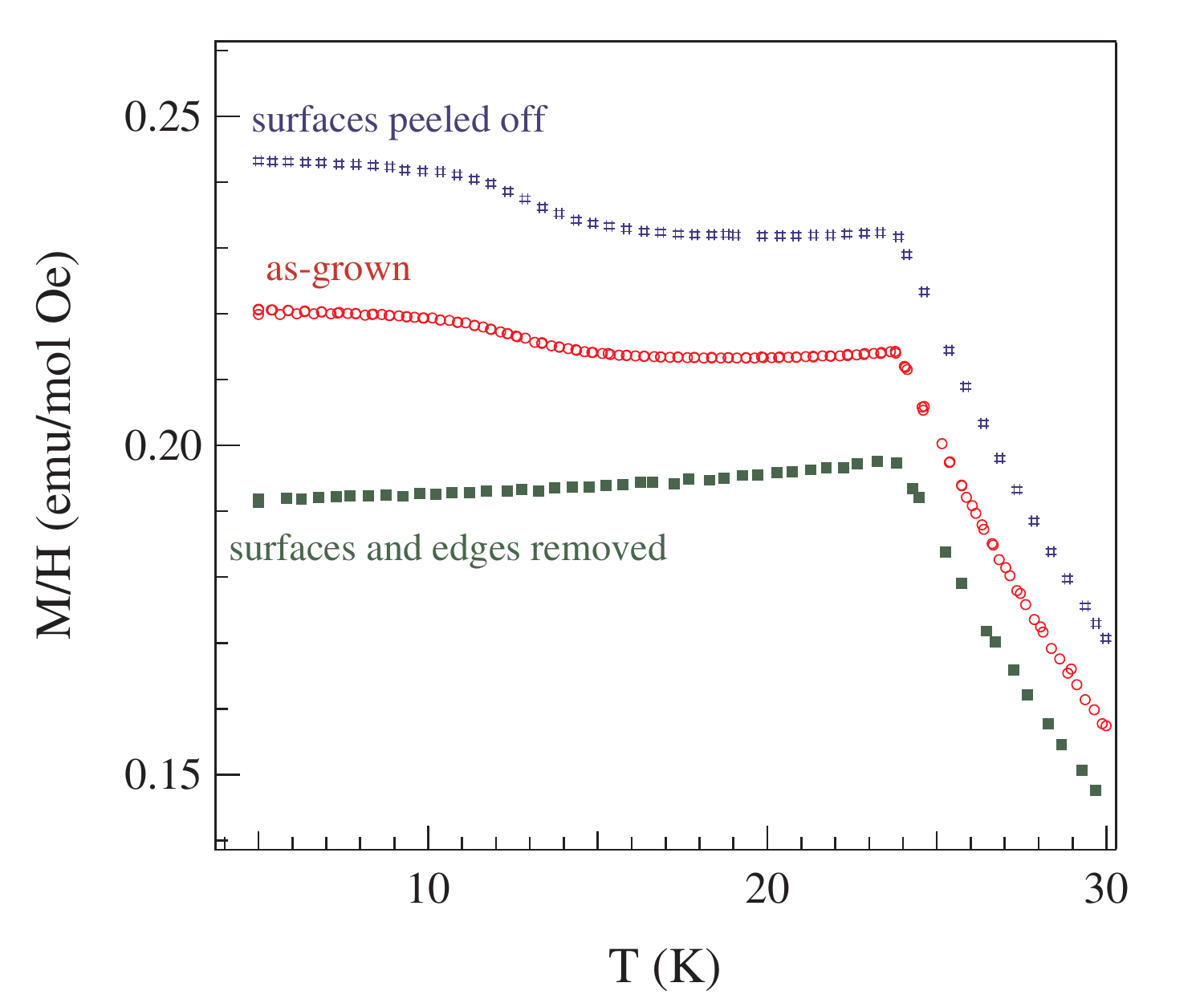}
\caption{(color online) Temperature dependence of magnetization measured in a magnetic field of 1\,kOe perpendicular to the crystallographic $c$-axis. The curves are shifted for clarity. The ferromagnetic signal disappears after cutting off the edges where the magnetic impurity Bi$_{2-x}$Mn$_x$Te$_3$ tends to stay.}
\label{Contamination}
\end{figure}

\begin{figure} \centering \includegraphics [width = 0.47\textwidth] {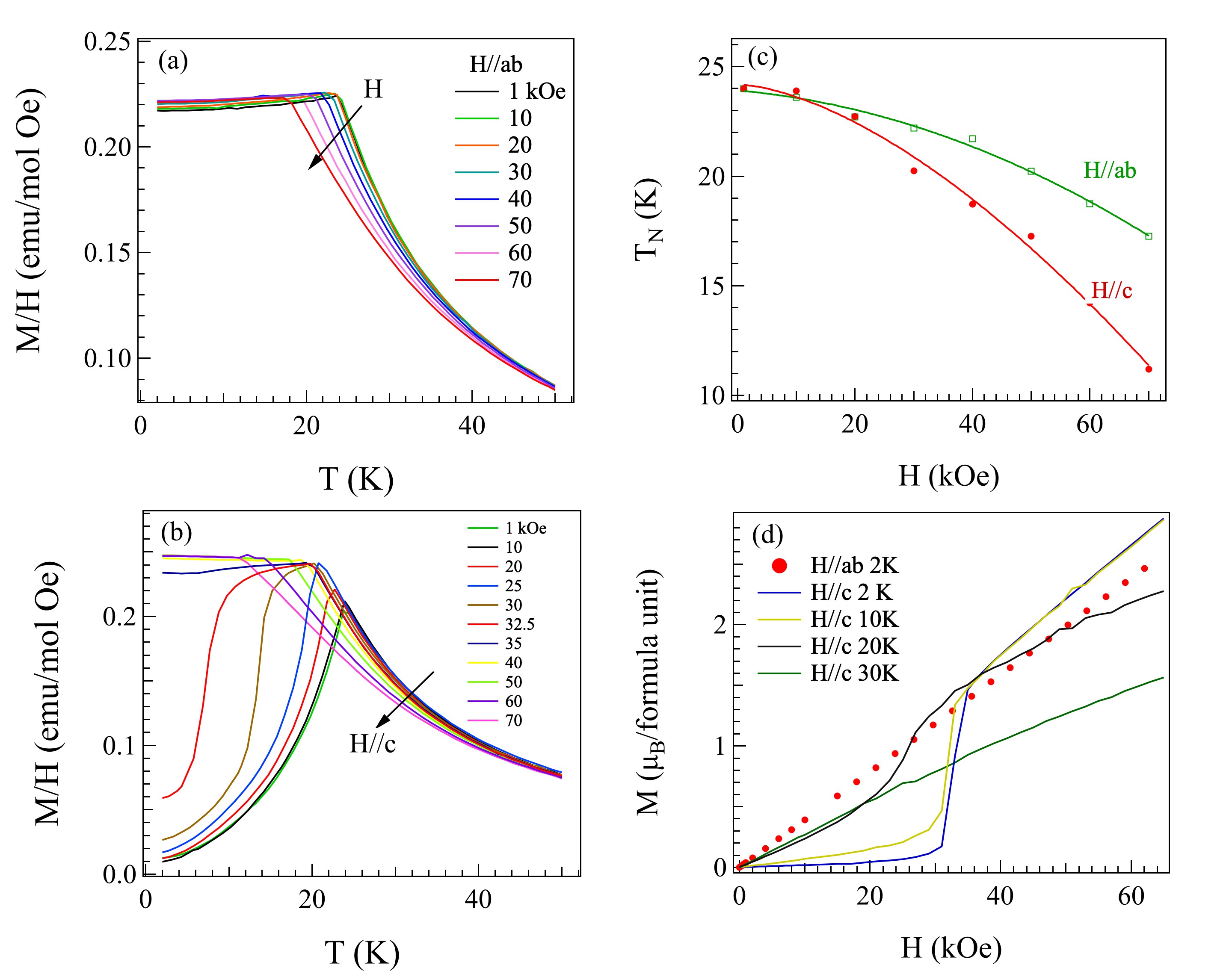}
\caption{(color online) (a,b) Temperature dependence of magnetization in various magnetic fields up to 70\,kOe perpendicular (H//ab)  and parallel (H//c) to the crystallographic $c$ axis, respectively. (c) Suppression of T$_N$ with increasing magnetic fields.  It should be noted that a metamagnetic transition occurs around 35\,kOe when the magnetic field is applied along the crystallographic \textit{c}-axis. The solid curves are a guide to the eye. (d) The field dependence of magnetization at different temperatures.}
\label{Mag}
\end{figure}

\section{Results and discussion}

\subsection{Magnetic and transport properties}
As reported previously, \cite{zeugner2018chemical,lee2013crystal} MnBi$_2$Te$_4$ can be synthesized only in a narrow temperature window around 600$^\circ$C, which makes the crystal growth rather challenging. Our growth strategy takes advantage of the low melting temperature of Bi-Te mixture. The melting tempertaure of Bi$_2$Te$_3$ is 585$^\circ$C, which is in the temperature range where MnBi$_2$Te$_4$ is stable and makes possible the crystal separation from flux by decanting. One concern of this growth strategy is that Bi$_{2-x}$Mn$_x$Te$_3$ melt might stay on the surface of the crystal and contribute a weak ferromagnetic signal at low temperatures. Fig.\,\ref{Contamination} shows the temperature dependence of the magnetic susceptibility of one MnBi$_2$Te$_4$ crystal cleaned differently as described below. The measurement was performed in a field of 1\,kOe applied perpendicular to the crystallographic \textit{c}-axis. For the as-grown crystal, there is a weak ferromagnetic signal below Tc$\sim$13\,K, which coincides with the ferromagnetic ordering temperature of Bi$_{2-x}$Mn$_x$Te$_3$.\cite{hor2010development} The measurement was then  performed on the same piece of crystal  after peeling off the surface layers. The presence of the low temperature ferromagnetism suggests negiligible amount of Bi$_{2-x}$Mn$_x$Te$_3$ on the surface of MnBi$_2$Te$_4$ crystals. We further cleaned the crystals by cutting off the edges using a sharp surgical blade. The nearly temperature independent magnetic susceptibility below T$_N$=24\,K suggests that the magnetic impurities of Bi$_{2-x}$Mn$_x$Te$_3$ tends to stay on the edges of the crystals. This is similar to the contamination of NdFeAsO single crystals by NdAs we observed before.\cite{yan2011contamination} Therefore, before magnetic measurements, we carefully cleaned the crystals by removing the edges using a surgical blade.

Figure\,\ref{Mag} (a, b) show the temperature dependence of the magnetic susceptibility measured in various magnetic fields applied perpendicular (labelled as H//ab) and parallel to the crystallographic \textit{c}-axis, respectively. The anisotropic temperature dependence agrees with previous report\cite{otrokov2018prediction} and suggests an antiferromagnetic order at T$_N$=24\,K. With increasing magnetic fields, the magnetic order is suppressed to lower temperatures. The suppression of T$_N$ with increasing magnetic fields is summarized in Fig.\,\ref{Mag}(c). A spin flop transition occurs around 35\,kOe when the field is applied along the crystallographic \textit{c}-axis. This is better illustrated by the M(H) curves shown in Fig.\,\ref{Mag}(d). A linear field dependence at all measured temperatures was observed when the magnetic field is applied perpendicular to the crystallographic \textit{c}-axis.  The data collected at 2\,K are shown as an example. When the magnetic field is applied parallel to the \textit{c}-axis,  a metamagnetic transition is observed when the field is larger than 35\,kOe. At 20\,K, the metamagnetic transition occurs in a wider field range around 25\,kOe. The metamagnetic transition disappears when the measurement is performed at a temperature above T$_N$. The observed metamagnetic transition suggests the magnetic moment is aligned along the crystallographic \textit{c}-axis, which agrees with the temperature dependence of the magnetic susceptibility and the magnetic structure revealed by neutron diffraction.

\begin{figure} \centering \includegraphics [width = 0.47\textwidth] {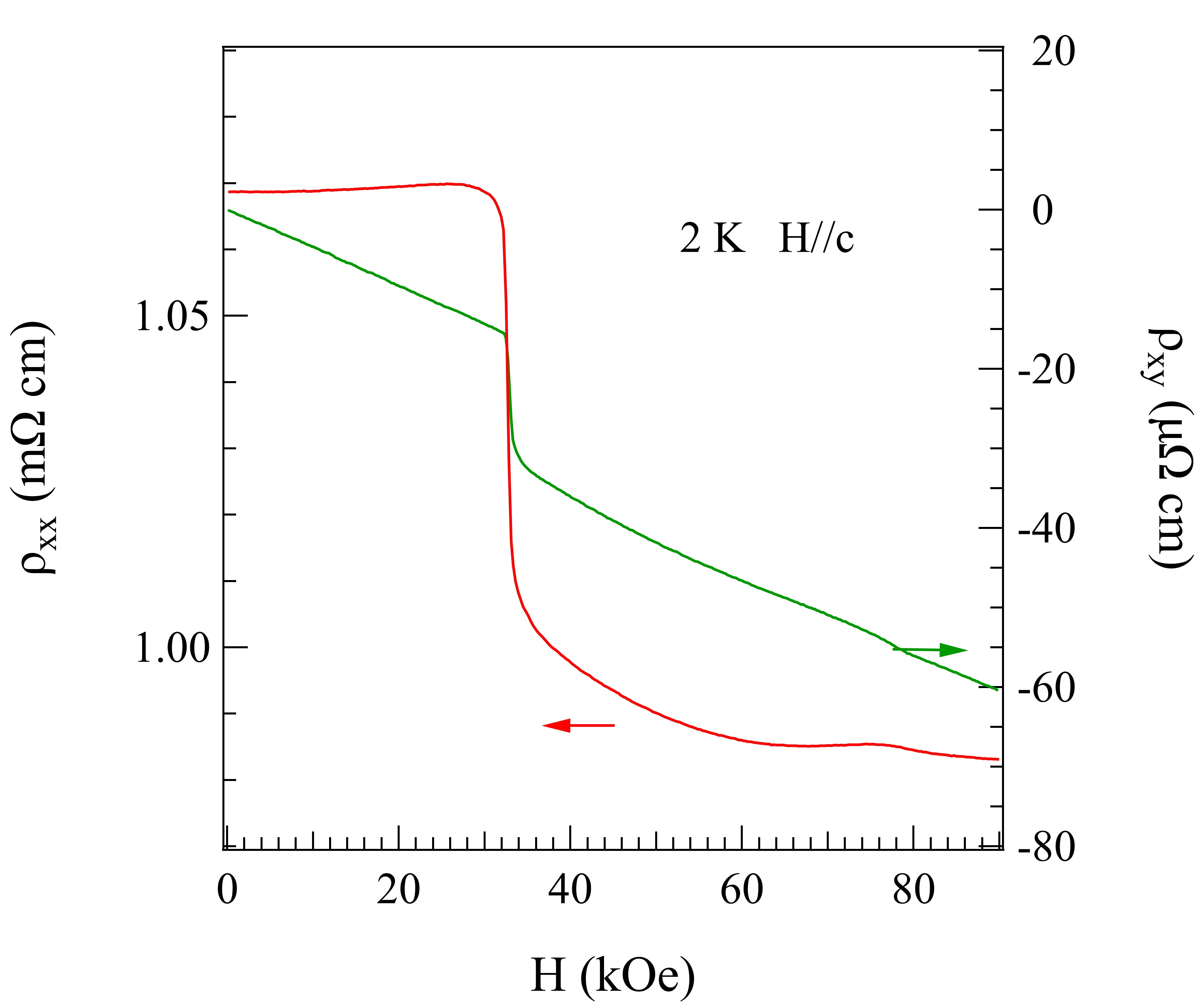}
\caption{(color online) Field dependence of resistivity at 2\,K. Around 32\,kOe where a metamagnetic transition occurs, a sharp drop was observed in both $\rho_{xx}$ and $\rho_{xy}$. A weak anomaly was observed around 78\,kOe above which the Mn spins are fully polarized. }
\label{RT}
\end{figure}

\begin{figure} \centering \includegraphics [width = 0.47\textwidth] {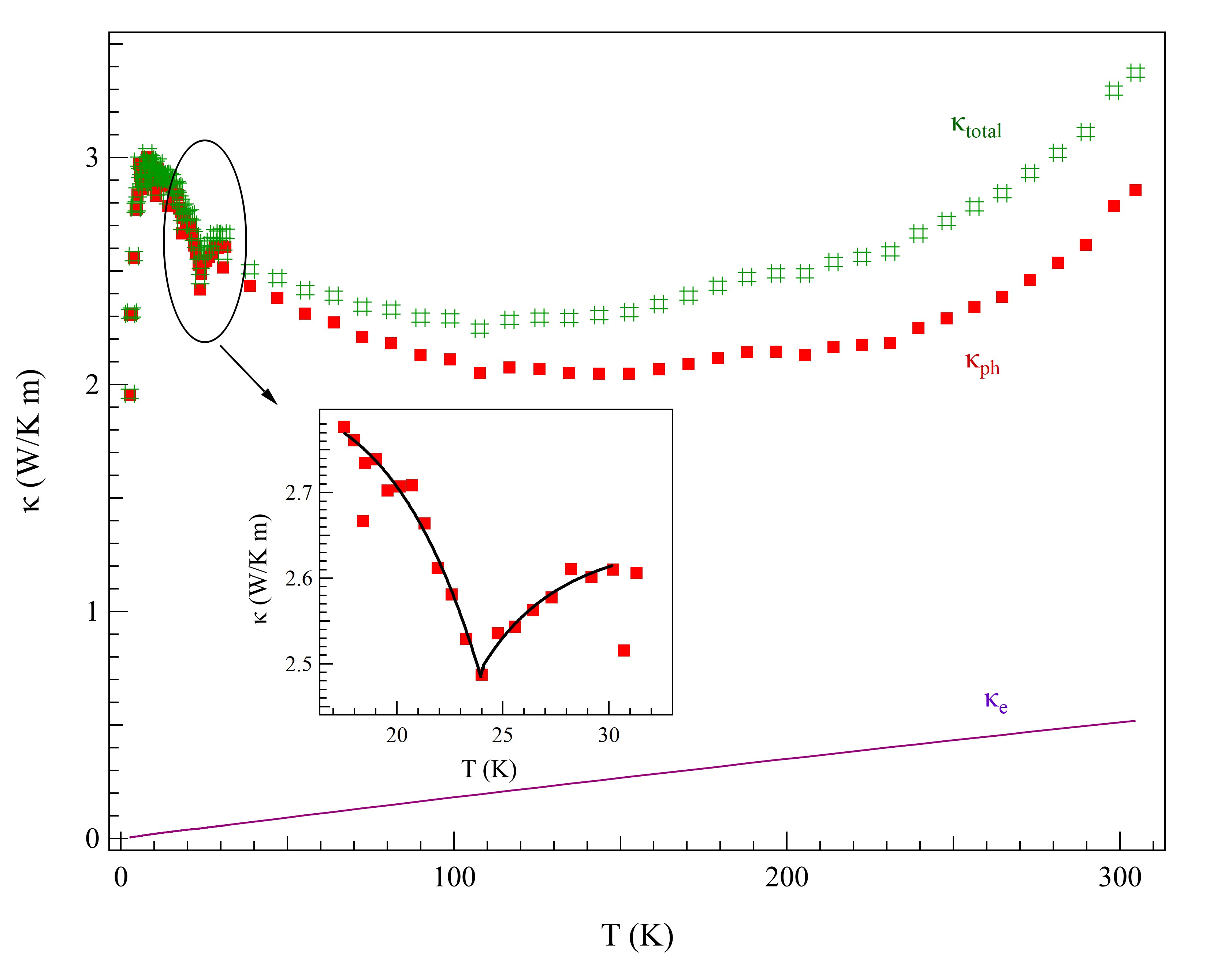}
\caption{(color online) Temperature dependence of thermal conductivity. The electronic thermal conductivity, $\kappa_e$, was estimated from the electrical resistivity data using the Wiedemann-Franz law. The lattice thermal conductivity, $\kappa_{ph}$, was obtained by subtracting $\kappa_e$ from the total thermal conductivity. Inset highlights the details of $\kappa_{ph}$ around T$_N$. The solid curves in the inset are a guide to the eye highlighting the critical scattering.}
\label{kappa}
\end{figure}

The temperature and field dependence of electrical resistivity was measured in the temperature range 2\,K$\leq$T$\leq$300\,K and in magnetic fields up to 90\,kOe.  The temperature and field dependence agrees with previous reports.\cite{lee2018spin,zeugner2018chemical,otrokov2018prediction} From the Hall coefficient at room temperature, the electron density is about 5.3$\times$10$^{20}$cm$^{-3}$ assuming one carrier band. Figure\,\ref{RT} shows the field dependence of in-plane electrical and Hall resistivity at 2\,K with the magnetic field applied parallel to the crystallographic \textit{c}-axis.  The electrical and Hall resistivity drops sharply around 32\,kOe where the metamagnetic transition occurs. Around 78\,kOe, a weak anomaly was observed in both curves where the Mn spins are fully polarized. The critical fields and large anomalous Hall effect agree with those reported by Lee et al.\cite{lee2018spin}

Figure \ref{kappa} shows the temperature dependence of the thermal conductivity, $\kappa$(T), in the temperature range 2\,K$\leq$T$\leq$300\,K. The thermal conductivity is low in the whole temperature range and weakly temperature dependent. A room temperature value of $\sim$3\,W/K m is comparable to that of a typical n-type of Bi$_2$Te$_3$. The low thermal conductivity signals strong scattering from electrons, magnetic fluctuations, and lattice defects. As presented later, our STM measurement found about 3\%
Mn$_{Bi}$ antisite defects, which might serve as an effective phonon scatterer due to the large mass difference between Mn and Bi. Without considering possible heat conduction by magnetic excitations, the lattice thermal conductivity, $\kappa_{ph}$, can be estimated by subtracting the electronic thermal conductivity from the total thermal conductivity. The electronic thermal conductivity, $\kappa_e$, can be estimated from the electrical resistivity data assuming the Wiedemann-Franz law is valid: $\kappa_e$=LT/$\rho$, where L is the Lorenz constant taken to be equal to 2.44$\times$10$^{-8}$\,V$^2$/K$^2$, T is the absolute temperature, and $\rho$ is the electrical resistivity.  $\kappa_e$ is small and decreases while cooling in the whole temperature range studied. $\kappa_{ph}$ follows the temperature dependence of the total thermal conductivity. A critical scattering behavior is observed around T$_N$=24\,K. The dip-like feature is highlighted in the inset of Fig.\ref{kappa}. Thermal conductivity studies of antiferromagnetic insulators have demonstrated a critical scattering effect that induces a dip in  $\kappa$(T) with a minimum in the region of the magnetic transition.\cite{slack1961thermal,slack1958thermal,lewis1973thermal} NiO and CoO are two typical examples showing this critical scattering effect in the $\kappa$(T) curve. As pointed by Carruthers,\cite{carruthers1961theory}  the anomalous dip in $\kappa$(T) is not a general phenomenon for all antiferromagnets due to varied lattice dynamics in different materials. For example, the perovskite KCoF$_3$ shows a dip anomaly in $\kappa$(T) at T$_N$ whereas KMnF$_3$ shows a minimum at T$_N$ and a glass-like thermal conductivity above T$_N$.\cite{suemune1964thermal} However, the critical scattering illustrated in the inset of Fig.\,\ref{kappa} suggests that the dominant role of the spin system is as an additional phonon scattering mechanism in MnBi$_2$Te$_4$ and the spin-lattice coupling is strong in MnBi$_2$Te$_4$.

Magnetic excitations can also carry heat, especially in low dimensional systems. A famous example is the spin-ladder compound Ca$_9$La$_5$Cu$_{24}$O$_{41}$ in which the thermal conductivity parallel to the ladder direction is nearly two orders of magnitude higher than the thermal conductivity perpendicular to the ladder direction.\cite{hess2001magnon} Similar magnetic heat transport has also been reported in (quasi-)2-dimensional materials such as La$_2$CuO$_4$.\cite{yan2003thermal} Considering the 2-dimensional arrangement of Mn-sublattice, heat transport by magnetic excitations is likely and this deserves further careful investigation.

\begin{figure} \centering \includegraphics [width = 0.47\textwidth] {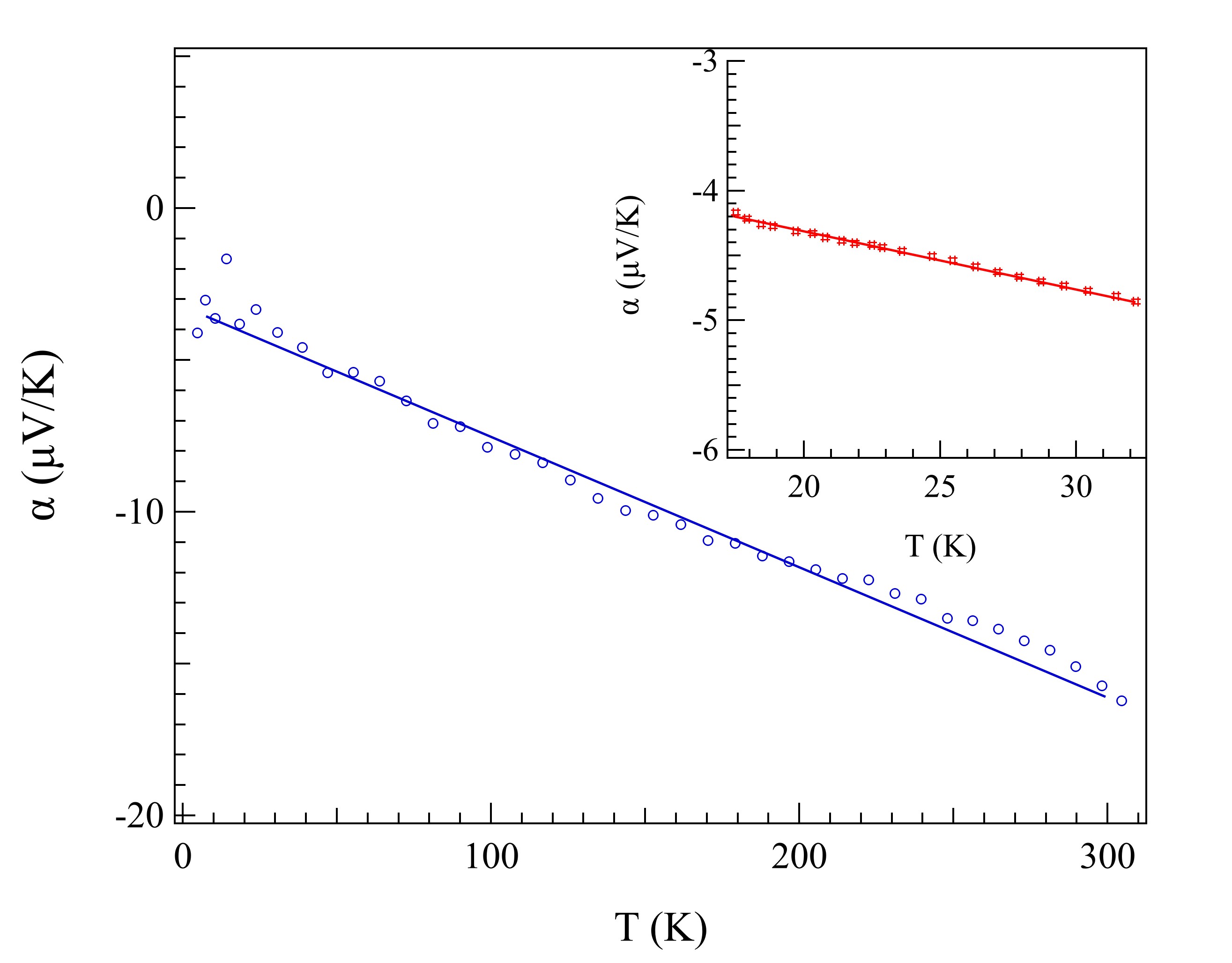}
\caption{(color online) Temperature dependence of thermopower. Inset shows a more careful measurement around T$_N$ which shows no anomaly.}
\label{Thermopower}
\end{figure}

Figure\,\ref{Thermopower} shows the evolution of the thermopower, $\alpha$(T), with temperature. At room temperature,  $\alpha$(T) has a value of -16$\mu$V/K. The negative sign of $\alpha$(T) signals electrons dominated charge transport and the small absolute value signals high electron concentration, consistent with the Hall data and STS result presented later. The absolute value of $\alpha$(T) decreases linearly upon cooling from room temperature to 2\,K. This linear temperature dependence corresponds to the characteristic diffusion thermopower of a metal. A careful measurement around T$_N$ (see inset of Figure\,\ref{Thermopower}) found no response of $\alpha$(T) to the magnetic order. $\alpha$(T) was also measured in magnetic fields up to 90\,kOe applied perpendicular to the crystallographic \textit{c}-axis. However, no magnetothermopower was observed in the temperature range 2\,K$\leq$T$\leq$80\,K. Thermopower is proportional to the logarithmic derivative of the density of states (DOS) with respective to energy at the Fermi level and it is sensitive to the asymmetry in the DOS near the Fermi level. The absence of any anomaly in $\alpha$(T) across T$_N$ suggests that the A-type antiferromagnetic order either does not modify the electronic band structure or the asymmetry in the DOS is maintained even though the band structure is changed across the magnetic order. For the latter case, magnetic ordering would have a larger effect on electrical resistivity beyond the reduction of spin-disorder scattering. It would be interesting to measure thermal conductivity and thermopower in magnetic fields parallel to the \textit{c}-axis to probe the effects of canted magnetism on the bulk properties.

\subsection{STM/STS}

\begin{figure} \centering \includegraphics [width = 0.47\textwidth] {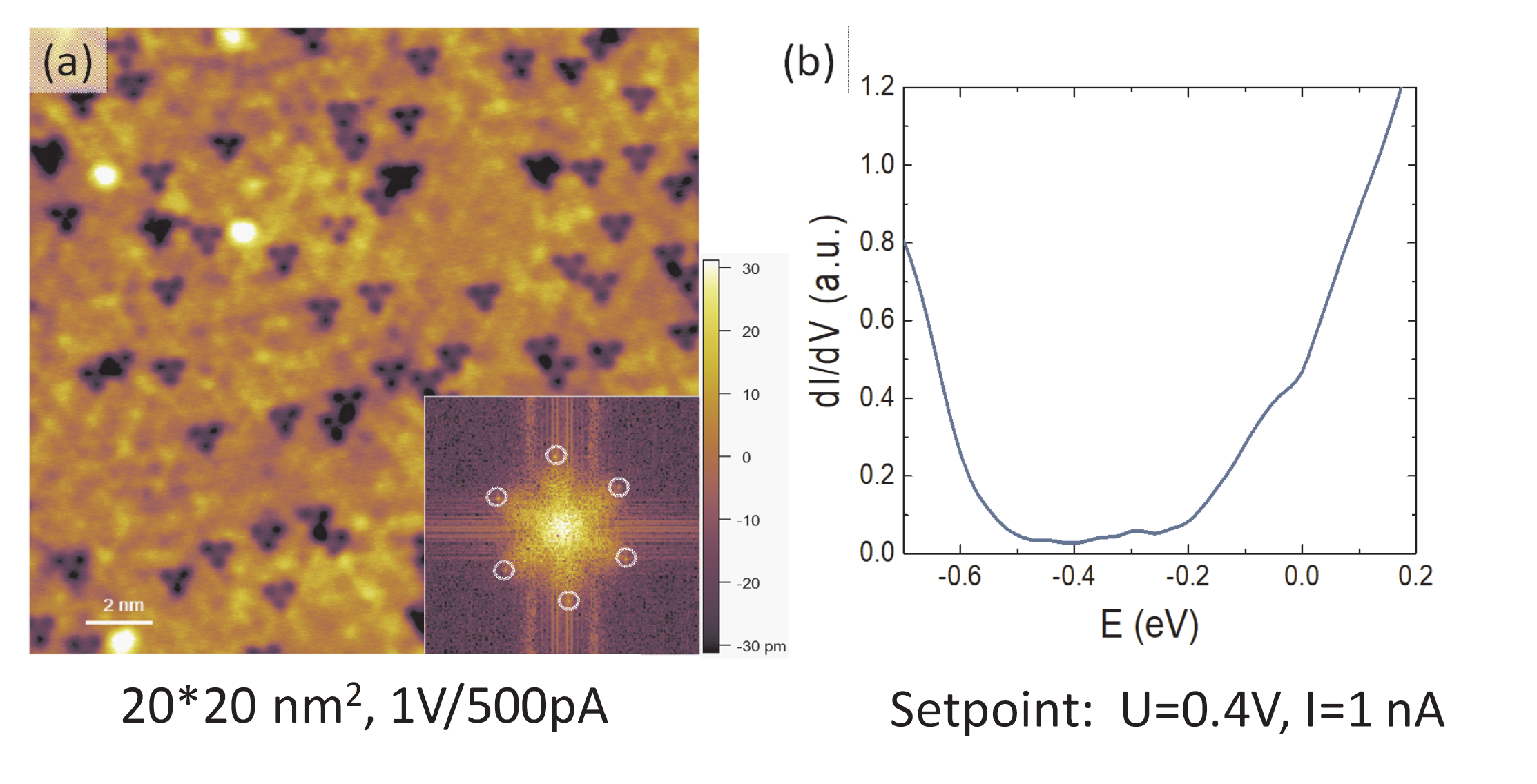}
\caption{(color online) (a) the STM image of the cleaved MnBi$_2$Te$_4$ terminating with the Te surface. Inset shows the Fourier transformation of the STM image. (b) The local density of states (LDOS) is measured by the spatially averaged conductance spectrum.}
\label{STM-1}
\end{figure}

Fig.\,\ref{STM-1} shows an STM image of a cleaved MnBi$_2$Te$_4$ single crystal terminating with the Te surface. Two types of defects can be observed on the surface: bright circular protrusions and dark clover-shape depressions. Presumably, they are respectively Bi$_{Te}$ antisites in the first layer and Mn occupying Bi sites (Mn$_{Bi}$) in the second layer, as assigned by the previous STM work on topological insulator Bi$_2$Se$_3$ \cite{dai2016toward} and Mn-doped Bi$_2$Te$_3$ \cite{hor2010development}. By counting the number of Mn$_{Bi}$ defects, it is estimated that Mn occupies about 3\% of the Bi sites in the second layer. In Mn-doped Bi$_2$Te$_3$, 1\% of Mn doping is sufficient to generate ferromagnetism.\cite{lee2014ferromagnetism} However, we did not notice any anomaly in the temperature dependence of magnetic susceptibility of a well cleaned crystal. The hexagonal Bragg peaks in the Fourier transformation of the STM image reveal the hexagonal lattice formed by the Te atoms, and from which the lattice constant is estimated to be 4.3${\AA}$.  The local density of states (LDOS) is measured by the spatially averaged conductance spectrum.  The valence band maximum (VBM) and the conduction band minimum (CBM) locate at around -0.5 and -0.2 eV, respectively. This is consistent with the recent APRES results [4]. The finite LDOS inside the band gap indicates possible existence of topological surface states.

\subsection{Neutron diffraction}

\begin{figure}
\centering \includegraphics[width=1\linewidth]{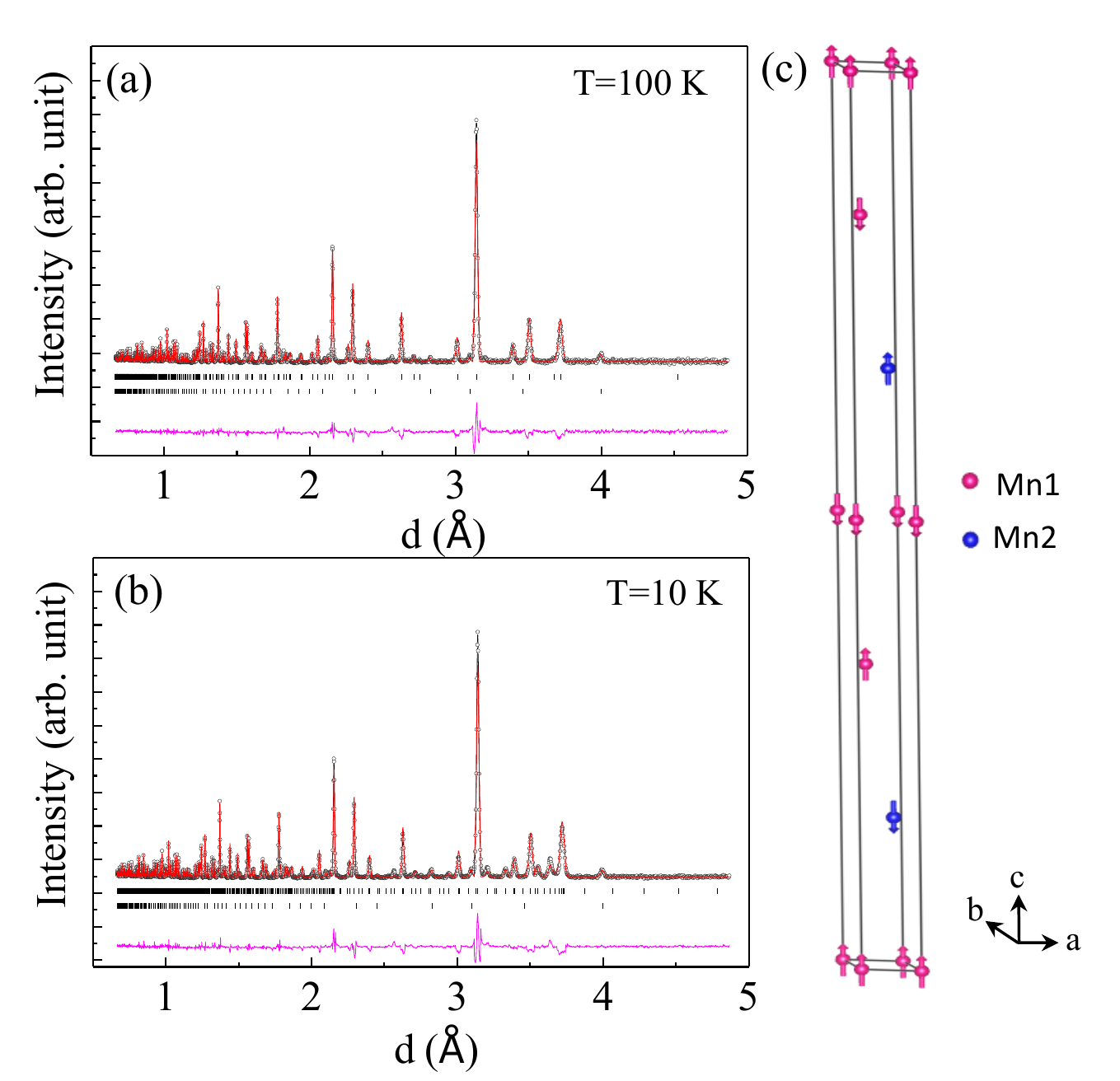} \caption{
(color online) Rietveld refinement fits to neutron diffraction patterns of MnBi$_{2}$Te$_{4}$ at (a) 100 K, and (b) 10 K. The observed
data and the fit are indicated by the open circles and solid lines, respectively. The difference curve is shown at the bottom.
The vertical bars mark the positions of Bragg peaks (nuclear peaks) in (a);
both nuclear and magnetic peaks in (b)) for MnBi$_{2}$Te$_{4}$ (top) and
impurity phase MnTe$_{2}$ (bottom). (c) The determined magnetic structure of MnBi$_{2}$Te$_{4}$, with two
coordinates Mn1 (0,0,0) and Mn2 (0.6667,0.3333,0.1667) in one magnetic cell.  }
\label{fig:Neutron}
\end{figure}

In order to determine the magnetic structure and ordered moment, we first performed neutron powder diffraction. The diffraction patterns at 100 K and 10 K are shown in Fig.\,\ref{fig:Neutron} (a) and
(b), respectively. Rietveld analysis confirms the trigonal structure with space group \textit{R-3m}
 (No.\,166), consistent with previous report.\cite{lee2013crystal} About 5\%wt MnTe$_{2}$  was identified to exist in the sample.
Our neutron diffraction results show no change in the crystal structure of this compound
 down to 10 K. The refined atomic positions and lattice constants at
 100 K and 10 K are summarized in Table I.

At 10\,K (see Fig.\,\ref{fig:Neutron}(b)), neutron diffraction observed some additional reflections that are absent at 100\,K and are of magnetic origin. These magnetic reflections can be indexed with a propogation vector \textit{\textbf{k}} =(0,0,1/2). Symmetry-allowed magnetic space groups are analyzed by the Bilbao crystallographic server to create
the PCR file for refinement. We used one magnetic unit cell ($a\times b \times 2c$) to refine both nuclear
and magnetic peaks to obtain the lattice information and magnetic structure
simultaneously. The refinement confirms an A-type antiferromagnetic order consisting of ferromagnetic layers coupled antiferromagnetically along the \textit{c}-axis with the magnetic space group P$_c$-3c1 (No.\,165.96). Refinements at 10\,K find an ordered moment of 4.04(13)$\mu_{B}$/Mn that is aligned along the crystallographic \textit{c}-axis. The determined magnetic structure is displayed in Fig. \ \ref{fig:Neutron} (c), consistent with previous theoretical predictions by density functional theory.\cite{eremeev2017competing, otrokov2018prediction} It is worth mentioning that symmetry analysis of the magnetic cell allows for different ordered moments at the two Mn sites at (0,0,0) and (0.6667,0.3333,0.1667), although our refinement suggests the same ordered moment. We also considered other possible magnetic structures, for example, the AFM order with the up-up-down-down stacking of ferromagnetic planes along the \textit{c} axis, or A-type AFM order with moments in \textit{ab}-plane, or G-type AFM order. None of these models provides reasonable refinement.

\begin{table}
\caption{Refined atomic positions and lattice constants at
T\,=\,100\,K and  10\,K for MnBi$_{2}$Te$_{4}$ with space group $R-3m$ (No.\,166).
Mn: 3\textit{a }(0, 0, 0); Bi: 6\textit{c} (0, 0, z); Te1: 6\textit{c} (0, 0, z); Te2: 6\textit{c} (0, 0, z)
  }
 \label{tab:lattice}
\begin{tabular} {llllll}
 \hline\hline
T &   Atom& Atomic position & \textit{a}(\AA{})& \textit{c}(\AA{}) &
   \\
\hline
100 K&   Bi  &   z= 0.4245(6)& 4.314(6) & 40.741(4) \\
 &   Te1 &    z= 0.1332(4)     &   &  &  \\
  &   Te2 &    z= 0.2940(8)     &   &  &  \\
 10 K&   Bi  &   z= 0.4247(6)& 4.309(7)& 40.679(5) \\
 &   Te1 &    z= 0.1324(7)     &   &  &  \\
  &   Te2 &    z= 0.2943(8)     &   &  &  \\
 \hline\hline
\end{tabular}
\end{table}

\begin{figure}
\includegraphics[width=3. in]{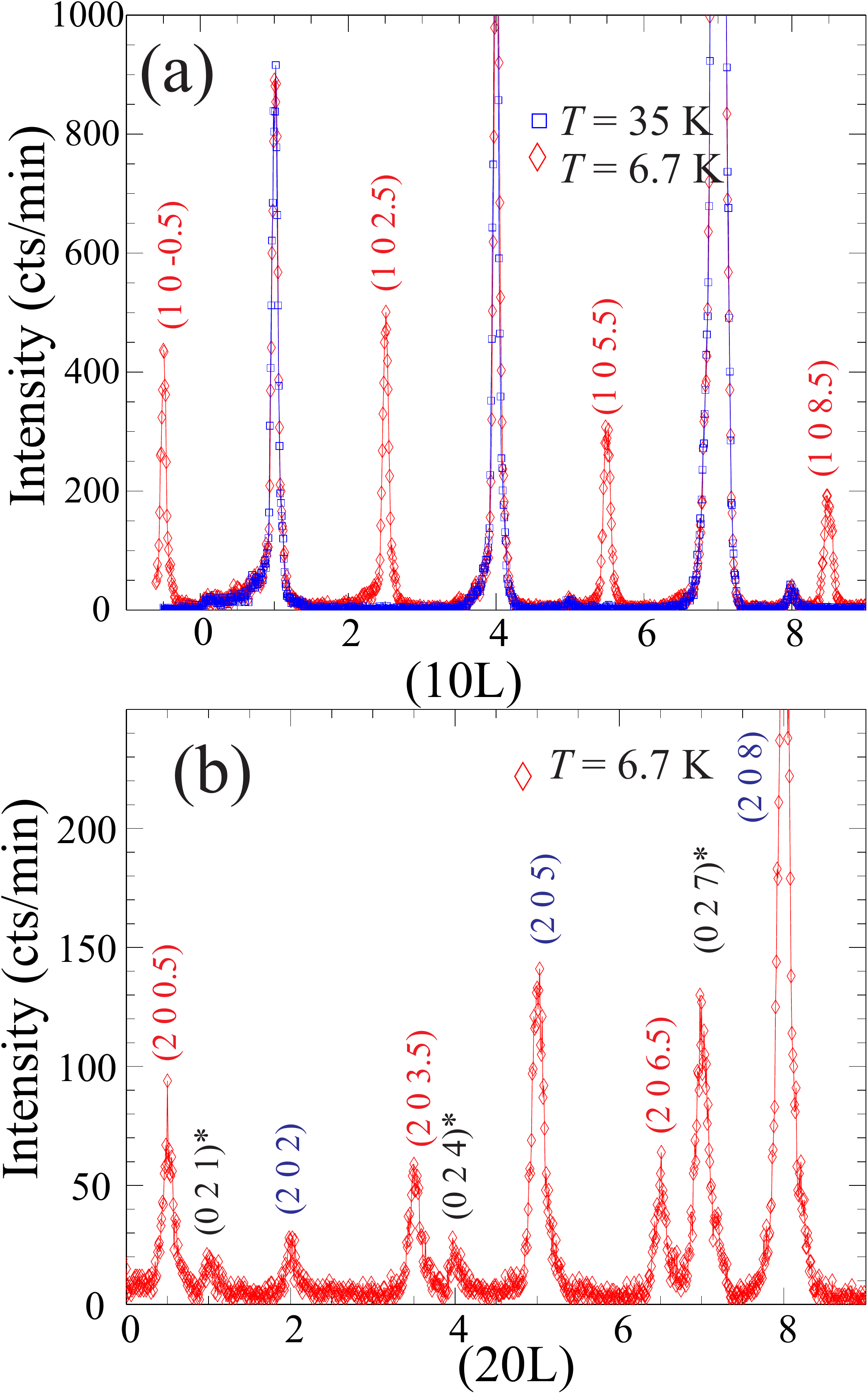}
\caption{(Color online)  (a) Neutron diffraction from single crystal along the (10L) above  (square symbols) and below (diamond symbols) T$_N$.   We note that peaks at the nominal (105) and (108) indicate that the crystal is twinned, namely  these peaks are (015) and (018) from the other domain. (b) (20L) scan at base temperature. Note the extra peaks from the other (02L) domain are marked with asterisk.}
\label{Fig:noL}
\end{figure}

\begin{figure}
\includegraphics[width=3.2 in]{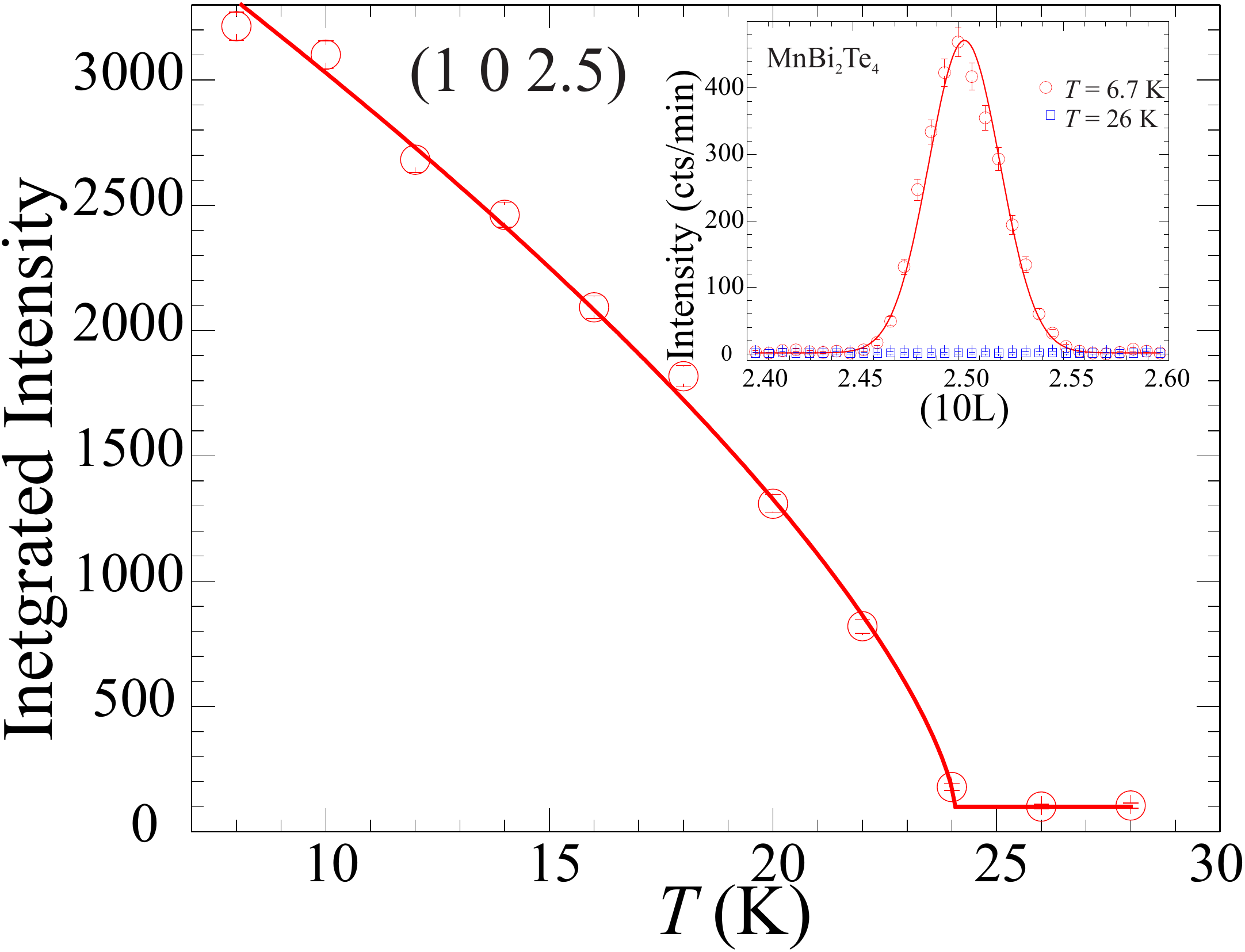}
\caption{(Color online) (left) Intensity versus temperature of the magnetic (1 0 2.5) including a fit to a power law $I \propto (1-T/T_N)^{2\beta}$ (solid line) that yields $T_{\rm N}= 24.1(2)$  K and $\beta = 0.35(2)$.   Inset shows intensity of magnetic Bragg reflections versus momentum transfer below and above $T_{\rm N}$. }
\label{Fig:OP}
\end{figure}

The magnetic structure shown in Fig.\,\ref{fig:Neutron}(c) is further confirmed by neutron single crystal diffraction measurements. Our measurements observed no weak reflections that could not be indexed by the A-type AFM structure described above. Figure\,\ref{Fig:noL}(a) shows diffraction patterns from the single crystal along the (10L) direction (using hexagonal indexing) above (square symbols) and below (diamond symbols) T$_N$ showing emerging half integer L magnetic Bragg reflections at base temperature, consistent with powder diffraction results. Similarly, Figure\,\ref{Fig:noL}(b) shows emerging half integer reflections along the (20L) reflections.  These half integer reflections and the absence of extra reflections along the (00L) at low temperature confirm doubling of the chemical unit cell due to the antiparallel arrangement of adjacent ferromagnetic Mn planes where the magnetic moment in each basal plane is along the \textit{c}-axis.
We note that peaks at the nominal (105) and (108) in Fig.\,\ref{Fig:noL}(a) indicate that the crystal is twinned, namely these peaks can be indexed as (015) and (018) reflections from the other domain. Consistent with that, the (20L) scans in Fig.\,\ref{Fig:noL}(b) also show extra peaks from the other domain at (021), (024), and (027) (marked with asterisk).  Figure\,\ref{Fig:OP} shows the temperature dependence of the integrated intensity of the magnetic (1 0 2.5) Bragg reflection.  A fit to a power law I$\propto$(1-T/T$_N$)$^{2\beta}$ yields T$_N$=24.1(2)\,K and $\beta$=0.35(2). The Neel temperature agrees well with that determined from magnetic and transport measurements.

\section{Summary}
In summary, we have successfully grown sizable single crystals of MnBi$_2$Te$_4$ out of a Bi-Te flux. The large crystals make possible the exploration and investigation of the intrinsic properties of MnBi$_2$Te$_4$ using various techniques including neutron diffraction and thermal conductivity measurements. Hall and STS measurements suggest the crystals are n-type with a carrier concentration of 5.3$\times$10$^{20}$cm$^{-3}$ at room temperature. MnBi$_2$Te$_4$ orders antiferromagnetically at T$_N$=24\,K. Our neutron powder and single crystal diffraction measurements confirm the proposed A-type antiferromagnetic order with ferromagnetic planes coupled antiferromagnetically along the \textit{c}-axis. The ordered moment is 4.04(13)$\mu_{B}$/Mn at 10\,K and aligned along the crystallographic \textit{c}-axis. The electrical resistivity drops upon cooling across T$_N$ due to the reduced scattering. The long range magnetic order also induces a critical scattering effect around T$_N$ in the temperature dependence of thermal conductivity. These changes suggest that the Mn spins are effective scatterers affecting the electrical and thermal transport. No anomaly in thermopower was observed across T$_N$, which indicates that the A-type antiferromagnetic order has negligible effect on the electronic band structure. However, the sharp change of electrical and Hall resistivity when going across the metamagnetic transition signals strong coupling between the canted magnetism and the bulk band structure. Fine tuning of the magnetism and/or electronic band structure is needed for the proposed topological properties of this compound. The growth protocol reported in this work provides a convenient route to high quality crystals where the electronic band structure and magnetism can be finely tuned by chemical substitutions.

\section{Acknowledgment}

Work at ORNL and Ames Laboratory was supported by the U.S. Department of Energy, Office of Science, Basic Energy Sciences, Materials Sciences and Engineering Division.  Ames Laboratory is operated for the U.S. Department of Energy by Iowa State University under Contract No. DE-AC02-07CH11358. The STM/STS work is supported by NSF grant DMR-1506618. A portion of this research used resources at Spallation Neutron Source, a DOE Office of Science User Facility operated by the Oak Ridge National Laboratory.

 This manuscript has been authored by UT-Battelle, LLC, under Contract No.
DE-AC0500OR22725 with the U.S. Department of Energy. The United States
Government retains and the publisher, by accepting the article for publication,
acknowledges that the United States Government retains a non-exclusive, paid-up,
irrevocable, world-wide license to publish or reproduce the published form of this
manuscript, or allow others to do so, for the United States Government purposes.
The Department of Energy will provide public access to these results of federally
sponsored research in accordance with the DOE Public Access Plan (http://energy.gov/
downloads/doe-public-access-plan).

\section{references}
\bibliographystyle{apsrev4-1}

%

\end{document}